# On the electronic structure of the charge-ordered phase in epitaxial and polycrystalline La$_{1-x}$Ca$_x$MnO$_3$ ($x$ = 0.55, 0.67) perovskite manganites

Rainer Schmidt

*University of Cambridge, Department of Materials Science and Metallurgy, Pembroke Street, Cambridge CB2 3QZ (UK); Now @: The University of Sheffield, Department of Engineering Materials, Mappin Street, Sheffield S1 3JD (UK);*

*Author's Email: rainerxschmidt@googlemail.com*

**Abstract** In this work the charge transport properties of charge ordered (CO) La$_{1-x}$Ca$_x$MnO$_3$ (LCMO) ($x$= 0.55, 0.67) epitaxial thin films and polycrystals are discussed following the recent controversy of localised electron states vs. weakly or de-localised charge–density wave (CDW) states in CO manganites. The transport properties were investigated by current vs. voltage, direct current resistivity vs. temperature, local activation energy vs. temperature, magnetoresistance and admittance spectroscopy measurements, which all indicated a localised electronic structure in the single CO phase. Delocalised charge anomalies observed previously may be restricted to phase separated materials.

Keywords: Charge-Ordering, Perovskite Manganites, Localised Electrons

PACS Numbers: 75.47.Lx; 71.23.An; 71.27.+a

# I. INTRODUCTION

Charge ordering (CO) phenomena in solid state materials are most commonly thought of as the alternating arrangement of localised cationic charge of different valence in a crystal lattice. CO has been studied most extensively in oxides: well known examples are the superconducting cuprates [1] and manganite perovskites.[2,3] Among the manganites, $Pr_{1-x}Ca_xMnO_3$ (PCMO) and $La_{1-x}Ca_xMnO_3$ (LCMO) have attracted considerable interest.[4] PCMO exhibits a meta-stable electronically insulating CO phase for $0.3 < x < 0.5$, which can be molten into a metallic phase by magnetic fields,[5] pressure,[6] electrical currents[7] or X-Ray radiation.[8] In LCMO a stable CO phase is present for $0.5 < x < 0.875$, which makes this system suitable for studying fundamental aspects of CO. The well known LCMO phase diagram and a typical charge and orbital ordering pattern are shown in Fig.1a,b.

The $La_{1-x}Ca_xMnO_3$ system has attracted considerable interest first during the early 1990s due to the discovery of colossal magneto-resistance (CMR).[9] More recently the interest in LCMO and related compounds is driven by the discovery of phase coexistence on different length scales and CO.[10-14] Such phase coexistence underlies the CMR effect and comes about due to the competition of different ordering mechanisms and exchange interactions between the valence d-electrons. If the competition for predominance is close (in terms of energy gain), phase coexistence can be observed over significant temperature and compositional ranges in one single crystal. Small local variations in the crystal field potential, stoichiometry or strain then result in phase separation.

Although, at first sight, charge-ordering features in solid matter seem to directly point towards full localisation of electrons and alternating arrangements of such localised charges, recent studies on perovskite CO manganite systems have disputed this conventional model [15] and have postulated a weakly or de-localised Charge –Density Wave (CDW) picture.[16, 17]

This new view of CO mechanisms developed from findings of commensurate and incommensurate charge periodicities in CO LCMO at $x \geq 0.5$,[16-18] and a weak commensurate phase for $x < 0.5$.[19] The periodicities of such charge modulations (along the *a* crystallographic orientation) were shown to differ from the original lattice periodicity and the modulation wave vector *q* was found to be correlated to the doping level *x* below the CO transition temperature $T_{CO}$ by $q \approx (1-x) \cdot a^*$, where $a^*$ is the reciprocal space vector of the parent orthorhombic P*nma* unit cell.[20] Above $T_{CO}$, *q* was reported to fall below $(1-x) \cdot a^*$ (Fig.1c).[21]

At first, these unusual superlattice charge modulations were explained in the conventional localised electron picture as an average of different temperature dependent $Mn^{3+}/Mn^{4+}$ plane stacking periodicities.[20] However, in a more recent electron diffraction study this conventional interpretation was claimed to be insufficient.[17] In LCMO $x = 0.52$ the charge modulation was found to be uniform and to have a non-integer periodicity even by reducing the electron spot size to 3.6 nm, which is below the expected stacking fault distance implanted into the hypothesized *q* = $0.5 \cdot a^*$ $Mn^{3+}/Mn^{4+}$ plane stacking by doping to $x = 0.52$ (Fig.1d). The authors in this study (Ref.17) assumed that these unconventional charge modulations arise solely from the cationic valence electron structure, and in consequence, they claimed that the localised $Mn^{3+}/Mn^{4+}$ plane stacking or stripe models may be insufficient to explain these findings. Electronically "soft" phases with weak electron-lattice coupling have been suggested [17] and a phenomenological Ginzburg-Landau theory was postulated [16] to describe these phases in a delocalised charge-density wave (CDW) picture.[22, 23]

At this point, it has to be stated that any CO pattern could be regarded as some kind of CDW. However, the term CDW is conventionally used only for systems which undergo a Peierls transition by cooling below the Peierls transition temperature $T_{PE}$. A non-linear collective charge transport mode occurs below $T_{PE}$ above a critical applied electric field ($E > E_T$) when the CDW is unpinned (CDW sliding).[24-26] Above $T_{PE}$, CDW materials commonly exhibit low-dimensional metallic delocalised electron transport. By cooling below $T_{PE}$ an additional (larger) lattice periodicity is introduced into the crystal (as is observed in a CO pattern) (see Fig.1e). The key difference between a conventional localised CO pattern and a CDW is that in a CDW system the additional charge periodicity is introduced by the movement of the positively charged atomic cores, and in a conventional CO pattern by ordered valence charge. This is contradictory to the assumptions made in Ref.17 that superlattice periodicities in a CDW system arise from valence electrons, and is contradictory to the conclusion in this work that a CDW is present due to the incompatibility of valence charge and lattice periodicities.

In a metallic delocalised electron CDW system the introduction of a new (larger) lattice periodicity by cooling below $T_{PE}$ leads to a split of the Brillouin zone (see Fig.1e). The partially filled metallic electron band splits into one fully filled valence and an empty conduction band, and metallic charge transport disappears. The energy costs for atom core movements are over-compensated by the energy gain ½$\Delta E$ of the electrons near the new valence band edge.[26] The opened gap $\Delta E$ (CDW gap) allows thermally activated insulating or semiconducting charge transport of delocalised electrons to occur.

It can be stated that the occurrence of a high temperature delocalised electron metallic phase is a prerequisite for a CDW state to occur at lower temperatures (below $T_{PE}$).

In fact, it has been reported that in LCMO ($x > 0.5$) metallic conduction occurs at high temperatures and an optical gap opens up by cooling below $T_{CO}$,[27] which would both be consistent with the interpretation of the unusual superlattice periodicity as the formation of a CDW.

In order to shed more light on the controversy of conventional localised vs. new delocalised electron CDW models to describe CO patterns, the investigation of the charge transport behaviour may be a powerful tool. In a fully localised electron system, charge transport is expected to be by localised polaron hopping.[28] In contrast, in a CDW system charge transport is expected to be metallic above $T_{PE}$ and may be consistent with delocalised electron transport by thermal activation of electrons across the CDW gap below $T_{PE}$. Furthermore, by the application of electric fields above the CDW unpinning field ($E > E_T$), at temperatures below $T_{PE}$, the collective type CDW sliding mechanism is expected to manifest itself by strongly non-linear current vs voltage (*I-V*) characteristics.[25] In polycrystalline samples, the crystalline disorder may not allow for the collective CDW transport to fully develop and only a change of slope in the linear *I-V* curves may be expected at $E_T$.[26] Contrarily, in coherently strained epitaxial layers the high crystalline order may potentially allow CDW sliding to occur.

Therefore, the *I-V* characteristics, resistivity ($\rho$) vs. temperature curves ($\rho$-$T$), the magnetoresistance (MR), local activation energy vs. temperature and admittance spectroscopy data from CO polycrystalline and epitaxial LCMO of compositions $x = 0.55$ and $0.67$ were studied. It was anticipated to find either: A) typical CDW transport features, or B) signs of localised electron transport.

The compositions investigated, $x = 0.55$ and $0.67$, show a bulk CO temperature $T_{CO}$ of $\approx$ 220K and $\approx$ 260K respectively.[3] It has to be noted here that $T_{CO}$ represents a simplified single CO transition temperature determined from charge transport, crystallographic or magnetic properties.[3] In a more precise electron diffraction study, the CO transition in polycrystalline LCMO $x = 0.67$ was reported to occur gradually across a temperature window:[14] CO clusters form below $T^* \approx$ 280K and percolation occurs below $T_p \approx$ 235K. This implies that CO and non-CO phases coexist over a wide temperature range above and below $T_{CO}$ (Fig.1a,c).

## II. EXPERIMENTAL

### A. Thin film deposition

Epitaxial LCMO ($x = 0.55, 0.67$) films have been deposited by pulsed laser deposition (PLD) on 750°C pre-annealed orthorhombic (001) $NdGaO_3$ substrates from commercial targets (Praxair Specialty Ceramics USA, purity $\Delta x = \pm\, 0.01$, average grain size $\approx$ 1µm) using an excimer KrF laser (248 nm) at a deposition temperature of 750ºC with flowing $O_2$ pressure of 15.1 Pa, and 1 Hz repetition rate of the laser pulse with fluence of 1.6 J/cm$^2$.

Excellent film crystallinity was confirmed using X-Ray Diffraction (XRD) (Philips X'Pert high resolution XRD; wavelength 1.54 Å), where the (004) out-of plane film reflection showed a 0.011º full width half maximum (FWHM) rocking curve peak. Film thickness was determined from X-ray fringes in $\omega$-$2\theta$ scans with an uncertainty $\leq$ 10%. The bulk LCMO *a* and *b* lattice vectors exhibit a 0.85% / 2.0% ($x = 0.67$) and 0.24% / 1.44% ($x = 0.55$) mismatch compared to the $NdGaO_3$ substrate surface. The film surface roughness was determined to be below 1 nm by atomic force microscopy (AFM).

## B. Electrode deposition

Two sets of four parallel rectangular Ag electrodes for charge transport measurements were deposited onto the film surfaces using direct current (DC) sputtering, one set along the charge-modulated *a* and one in the non-modulated *b* crystal direction. DC sputtering leads to intimate adhesion of Ag particles on the film surface on an atomic scale, leading to highly ohmic contacts. Interface effects may be minimal, in contrast to painted electrodes with considerable air gaps between electrode and film, and imperfect contact between electrode particles and epitaxial layers of low surface roughness.

Polycrystalline bulk samples were obtained from the PLD bulk targets. Four rectangular Ag electrodes were deposited onto the pellet surfaces by thermal evaporation, again leading to intimate electrode - sample contact and ohmic behaviour.

## C. Charge transport measurements

A standard automated data acquisition system was used for 4-point current vs. voltage (*I-V*) measurements at 65 K – 573 K in films and polycrystals. Stepwise steady state voltage biased measurements were carried out at various temperatures under zero and $H = 0.5$ Tesla applied magnetic field. Steady state conditions imply that complete temperature and resistance stability was assured before taking *I-V* readings. At each temperature, an *I-V* curve was recorded and resistivity was determined from the slope in the strictly linear regime, the estimated current cross section and electrode distance. Magnetoresistance (MR) was calculated as $[\rho(H)-\rho(0)]/\rho(H)$.

For alternating current (AC) admittance spectroscopy measurements both sides of the polycrystalline pellets ($x = 0.55, 0.67$) were covered with Ag electrodes by thermal evaporation and covered with quick drying silver paint. A standard automated data acquisition system was used to perform Admittance Spectroscopy at 10 K using an Agilent E4980A LCR meter. The applied AC voltage signal had amplitude of 10 mV, which corresponded to ≈ 50 mV/cm applied electric field amplitude.

## III. RESULTS

### A. Current vs. voltage characteristics (*I-V*)

#### 1. Epitaxial thin film LCMO ($x = 0.55, 0.67$)

In epitaxial films linear charge transport was observed in the entire range up to the high voltage limit (15V/cm). Therefore, a potential critical CDW unpinning electric field $E_T > E_C$ would be at least two orders of magnitude larger than for standard transition metal oxide CDW systems. $E_T$ in single crystal $K_{0.3}MoO_3$ and $Rb_{0.3}MoO_3$ (blue bronzes) at 77K was quoted to be 113mV/cm [29] and ≈ 80mV/cm,[30] in $TaSe_3$ at 130K ≈ 300mV/cm [26] and in $NbSe_3$ at 26.5K 120mV/cm.[31] The thin film samples were tested for non-linearity in the low electric field limit (Fig 2a). Again, strict linearity was measured along the *a* charge modulated crystal direction down to ≈ 15 mV/cm in both $x = 0.55$ and $x = 0.67$ epitaxial films.

#### 2. Polycrystalline LCMO ($x = 0.55, 0.67$)

The *I-V* characteristics of the polycrystalline LCMO samples investigated showed deviations from linearity above a critical applied electric field $E_c$. Such a critical field may not be interpreted as a typical CDW unpinning field $E_T$, as will be argued in the following, and is therefore termed $E_c$. Fig.2b indicates that $E_c ≈ 2V/cm$ for $x = 0.67$ at 200 K and $E_c ≈ 10V/cm$ at 130 K. A possible interpretation of such non-linear transport as a typical non-linear CDW mechanism has been suggested for polycrystalline $Pr_{0.63}Ca_{0.37}MnO_3$.[32] Contrarily, it is suggested that the non-linearity observed here may be a result of self-heating of the sample, which is expected in an insulating material with a negative temperature coefficient of resistance.[33] Hysteretic current response from an applied voltage signal with a sine-wave amplitude is shown in Fig.2c. This is consistent with sample heating, whereas standard CDW systems do not show hysteretic *I-V* curves.[34, 35] The *I-V* curves measured were tested for non-linearity in the low electric field limit to search for unusually low $E_T$, but strict linearity for $x = 0.55$ and $x = 0.67$ polycrystals down to the resolution limit of $\approx 5$ mV/cm was found. Such potential $E_T$ in polycrystals would be unreasonably low, because extended grain boundary defects are expected to increase $E_T$.[26]

### B. Electrical resistivity

Resistivity vs. temperature ($\rho$-$T$) data was collected for LCMO ($x = 0.55$ and $x = 0.67$) epitaxial thin films and LCMO ($x = 0.55$ and $x = 0.67$) polycrystals. Data was plotted as $\ln(\rho)$ vs. $1/T$ (Fig.3).

#### 1. Epitaxial thin film LCMO ($x = 0.55$ and $x = 0.67$)

The $\ln(\rho)$ vs. $1/T$ curves showed clear bending (Fig.3) at $T < T_{CO}$ and followed a localised electron variable-range hopping (VRH) power law for small polarons,[28] typical of strongly correlated electron systems:

$$\rho = C\, T^{\alpha} \, \exp\left(\frac{T_0}{T}\right)^p \tag{1}$$

where $C$ is a constant, $T_0$ a characteristic temperature and $\alpha$ describes the pre-exponential temperature dependence.

The data in Fig.3 could be linearised by assuming VRH exponents of $p \approx 0.3 - 0.5$. This finding is a strong indication for a localised electron structure, because VRH models are valid only in localised electron hopping systems but not for delocalised electron transport.[36] The resistivity vs. temperature curves are fully consistent with a localised electron structure, but may not be explained in a delocalised electron picture.

## 2. Polycrystalline LCMO ($x = 0.55, 0.67$)

VRH was indicated in LCMO ($x = 0.67$) polycrystals with $p = 0.3$, but an exception was found in the data for LCMO ($x = 0.55$) although $x = 0.55$ epitaxial films showed transport behaviour consistent with VRH and the $x = 0.67$ samples. The results from $x = 0.55$ polycrystals may be influenced by the proximity of the doping composition $x = 0.55$ to the $x = 0.5$ compositional phase boundary, which separates metallic charge transport in $x < 0.5$ samples (below $T_C$) and insulating transport for $x > 0.5$. The $x = 0.55$ polycrystals, although macroscopically insulating, may contain small fractions of a metallic phase, which is commonly observed in phase coexistence materials such as the manganites. This is indicated by the horizontal (blue) lines in Fig.1a around the $x = 0.5$ phase boundary. In the presence of such typical phase coexistence in the proximity of the phase boundary, pure VRH may not be observable due to the electronic heterogeneity of the sample. This effect may not occur in epitaxial layers, because it was reported that the phase boundary can drop down to lower $x$ values due to epitaxial constraint.[37]

In CDW systems metallic conduction occurs above $T_{PE}$, and has indeed been reported in polycrystalline LCMO ($x > 0.5$) above $T > 425K$.[27] This metallic type conduction could not be reproduced in this study here, despite several attempts on all sample types and compositions investigated. The CDW typical metallic conduction at high temperature ($T > T_{PE}$) is missing.

### C. Magneto-resistance

Fig.4 displays a negative magnetoresistance (MR) for $x = 0.67$ polycrystals of decreasing magnitude with increasing temperature at low magnetic fields of $H = 0.5$ Tesla. In a low magnetic field approximation,[38] this is consistent with localized electron transport. In the $x = 0.55$ samples a peculiar positive MR effect at around $160K < T < 260K$ occurred. The low magnetic field limit of MR may be exceeded in this case, and positive MR is then expected for predominantly localized electron transport.[38]

A possible localised electron structure may be less evident in the MR behaviour, because negative MR occurs in localised and delocalised electron systems. However, delocalised electron systems usually exhibit negative MR only, and the occurrence of positive MR in one sample points towards a localised electron structure.

### D. Local activation energies

Steady-state resistivity vs. temperature measurements ($\rho$ vs. $T$) proved to be accurate with minimised scatter in $\ln(\rho)$ vs. $1/T$ for thin films and polycrystals, allowing differential analysis of the data. The gradient of $\ln(\rho)$ vs. $1/T$ curves was interpreted as a local activation energy for thermally activated charge transport and was plotted vs. $T$ in Fig.5. $\varepsilon_3$ values represent the activation energy at one specific temperature only, in contrast to global activation energies ($E_A$). The low noise level in the differentiated data curves in Fig.5 demonstrates the very high accuracy of resistivity determination by *I-V* curve measurements at steady state conditions.

A potential pre-exponential temperature dependence of resistivity $T^{\alpha}$ (equation 1) was neglected. All physically meaningful values $-3/2 < \alpha < 1$ were tested and lead to small quantitative changes only and the general $\varepsilon_3$ vs. *T* trends were unaffected. A change from negative to positive gradient of $\varepsilon_3$ vs. *T* upon cooling in Fig.5 clearly indicated the charge ordering transition, i.e. $\approx$ 225K for polycrystalline and epitaxial *x* = 0.67 samples, and $\approx$ 185K/ 235 K for polycrystalline/epitaxial *x* = 0.55 samples.

Additionally, the optical gap 2Δ for polycrystalline LCMO (*x* = 0.55 and *x* = 0.67) was plotted in Fig.5; data was taken from Kim et al.[27] It can be seen that in the charge-ordered regime ($T < T_{CO}$) the optical gap shows the opposite trend compared to the local transport activation energy $\varepsilon_3$. This $\varepsilon_3$ behaviour is inconsistent with thermally activated delocalised electron transport across the 2Δ gap. In a delocalised electron system, thermally activated insulating charge transport across a band gap would suggest that the charge transport activation energy may be compatible with the optical gap and shows a similar trend with temperature.

In *x* = 0.55 CO polycrystals at $T < T_{CO}$, $\varepsilon_3$ showed a stronger increase with temperature inconsistent with VRH. This behaviour in *x* = 0.55 polycrystals was explained by the proximity of the doping level to the *x* = 0.5 phase boundary in section III./B./2.

The trends of the local activation energies $\varepsilon_3$ vs. *T* are not compatible with the opening optical gap 2Δ. Therefore, a charge transport mechanism based on thermal activation of delocalised electrons across a CDW gap is not indicated. Indications for a CDW typical metallic conduction at high temperatures $T > T_{PE}$ are again missing.

In the magnetoresistance (MR) behaviour, a peak structure in MR vs. $T$ was found for $x = 0.67$ polycrystals near $T_{CO}$ (Fig.4), indicating a gradual CO transition across a certain temperature range. This peak structure also occurred in $\varepsilon_3$ vs. $T$ for the $x = 0.67$ polycrystal (Fig.5a). The transitional temperature ranges displayed correlate with the suggestion of a gradual formation of the CO phase in $x = 0.67$ polycrystals, and with the formation temperatures $T^*$ and $T_p$ reported previously.[14] The CO transition in the epitaxial $x = 0.67$ films was reflected as a more abrupt change in $\varepsilon_3$ vs. $T$ (Fig.5a), indicating a 1$^{st}$ order transition.

From Fig.5 it can be noted that the paramagnetic to anti-ferromagnetic transition at $T_N \approx 130K$ ($x = 0.67$) and $\approx 150K$ ($x = 0.55$)[3] is not reflected in the transport behaviour. It may therefore be concluded that charge transport exhibits no perceptible dependence on the random ($T > T_N$) or anti-parallel ($T < T_N$) d-electron spin orientation.

### E. Admittance Spectroscopy

Fig.6 shows admittance ($Y'$) spectra of polycrystalline LCMO ($x = 0.55, 0.67$) samples at 10K. At such low temperatures, the intrinsic bulk or any other high frequency ($f$) relaxation process are expected to be displayed in the spectra. $f$ independent admittance is found at low and intermediate $f$ ranges. At high $f$, considerable admittance dispersion occurs and is manifested in a constant slope of $Y'$ vs. $f$. This behaviour is consistent with a single high $f$ relaxation process and the data follows Jonscher's universal response law:[39]

$$\frac{Y'}{\sigma_{dc}} = 1 + A f^n \tag{2}$$

where $\sigma_{dc}$ represents the DC conductivity of the respective contribution and corresponds to the $f$ independent $Y'$ plateau, $A$ is a material specific constant. Good fits to the data are displayed in Fig.6, which were obtained using a numerical weighted least linear square routine to fit to equation (2).

No additional $Y'$ plateau is observed in the $Y'$ vs. $f$ curves in Fig.6 and only one high $f$ contribution may be present. This was confirmed by analysis of the complex plane impedance plots (Fig.7), where only one strongly suppressed semicircle can be observed for each sample. At high $f$, no additional relaxation process is indicated, because the data curves approach the origin of the graph. The strong suppression of the complex plane impedance semicircles can be interpreted as a broad dispersion of relaxation times,[40] which may indicate a high degree of disorder in the material. In CDW systems such as the blue bronze $K_{0.3}MoO_3$,[29] an additional high frequency relaxation process is expected. No such relaxation was found in CO LCMO here.

## IV. DISCUSSION

The $I$-$V$ curves of polycrystalline and thin film LCMO ($x = 0.55$ and $x = 0.67$) investigated did not indicate CDW transport in terms of non-linear conduction. In epitaxial thin films, charge transport is linear up to 15V/cm, and for CDW transport to occur, the CDW unpinning field $E_T$ would be unreasonably large. The possibility of an unusually low $E_T$ may be ruled out by the strict linearity of charge transport in polycrystals (down to 0.5 mV/cm).

Resistivity vs temperature measurements revealed that charge transport is insulating above and below $T_{CO}$. A high temperature metallic phase, which is a prerequisite for CDW states to form, was not detected in any of the samples investigated.

The admittance spectra measured on LCMO ($x = 0.67$) polycrystals followed Jonscher's universal response law, which is strongly indicative of high degrees of disorder and a localised electron structure.[41] Disorder favours electron localisation and Jonscher's universal response law is commonly observed in disordered systems exhibiting localised electron hopping but not in systems with delocalised electrons. Manganite perovskites are well known to display strong potential disorder,[42] in agreement with the findings presented here. A second high frequency relaxation process, which is expected in CDW systems, was not detected.

None of the charge transport properties investigated in polycrystalline and epitaxial $La_{1-x}Ca_xMnO_3$ ($x = 0.55, 0.67$) showed any evidence for contributions from a collective mode CDW type transport and all conform to the conventional localised electron model.[15] The only known metal oxide systems exhibiting CDW transport are the blue bronzes, which possess quasi-1-dimensional metallicity above $T_{PE}$ as a result of dimeric crystal symmetry. This symmetry does not occur in the manganites, making an equivalent CDW scenario unlikely.

In previous studies, delocalised electron features in CO manganites have been observed in compositions close to the phase boundary (i.e. $x = 0.5 - 0.52$),[17, 43, 44] where phase coexistence of CO and metallic phases can be expected at low temperatures. The observation of delocalised charge anomalies in these materials may not be a surprise regarding the fact that a metallic phase may be present. The formation of a CDW state could still potentially be possible though and arise from the metallic phase in LCMO. However, such CDW's needed to be associated to the metallic phase and have nothing in common with charge ordering except the superlattice electron diffraction spots as a manifestation of an extra charge periodicity.

In single phase charge ordered LCMO in the absence of metallic phases ($x \geq 0.67$), the conventional charge ordering stripe model in manganites may be a conclusive reflection of the localised character of charge, in contrast to recent claims.[17] The temperature dependent charge periodicities above $T_p$ shown in Fig.1c may well be ascribed to phase coexistence as well, in this case of insulating CO and non-CO phases, as represented by the vertical (green) lines in Fig.1a at the PM – CO phase transition.

## V. CONCLUSIONS

Although the formation of the CO phase resembles the formation of a CDW state, clear evidence for CDWs from a collective charge transport mode is missing in highly crystalline epitaxial thin films LCMO ($x = 0.55$ and $x = 0.67$), and in polycrystalline $x = 0.55$ and $x = 0.67$ samples. The CDW analogy in manganites may hold only for the unusual charge periodicities observed in electron diffraction patterns at compositions where phase coexistence of metallic and non-metallic phases is likely. The electronic structure of the CO single phase in manganites may be regarded as localised. No evidence for CDW features has been found in single phase CO perovskite $La_{1-x}Ca_xMnO_3$ ($x > 0.5$) systems.

## ACKNOWLEDGMENTS

This work was supported by the Leverhulme Trust. The author wishes to thank Prof Ulrich Weiss, Prof Ortwin Hess and Prof Paul Midgley for useful discussions. Thanks to Dr Gavin Burnell, Dr Diana Sanchez and Dr James Loudon for the help provided. Thanks to Prof Derek Sinclair for his help.

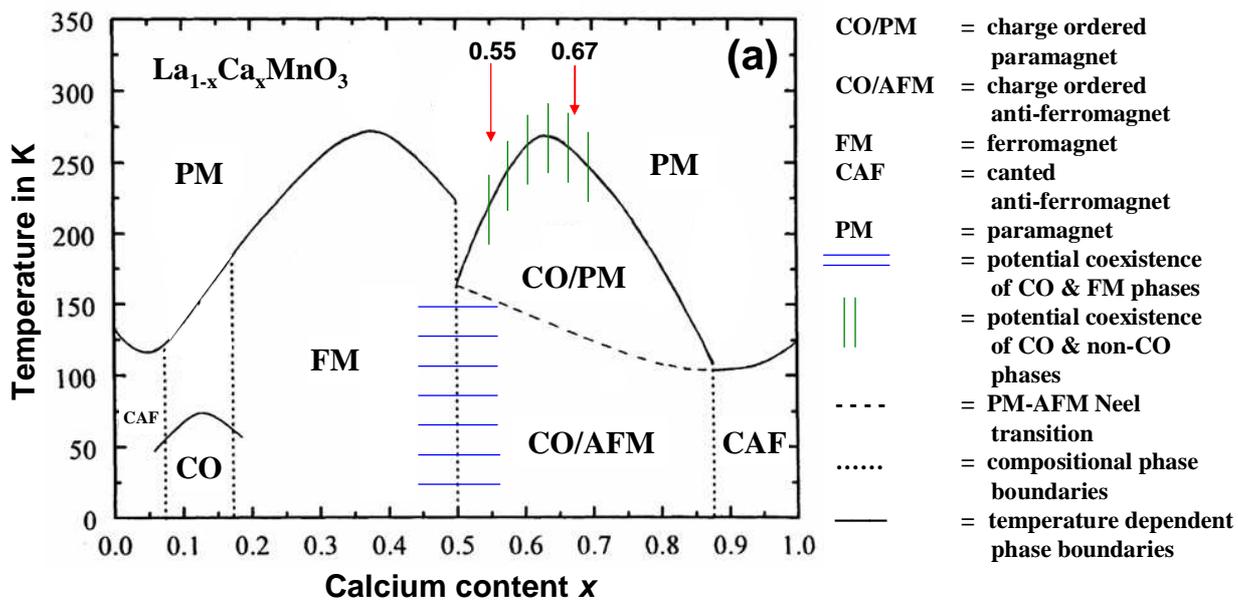

**Fig 1 (a)** (Colour Online) Phase diagram of the La$_{1-x}$Ca$_x$MnO$_3$ system (reproduced from [3])

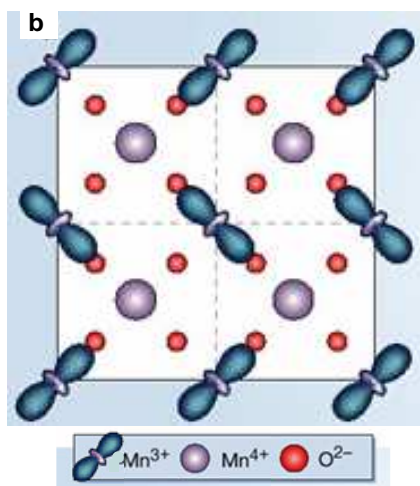

**Fig 1 (b)** (Colour Online) Charge and orbital ordering pattern in La$_{0.5}$Ca$_{0.5}$MnO$_3$ (reproduced from [4])

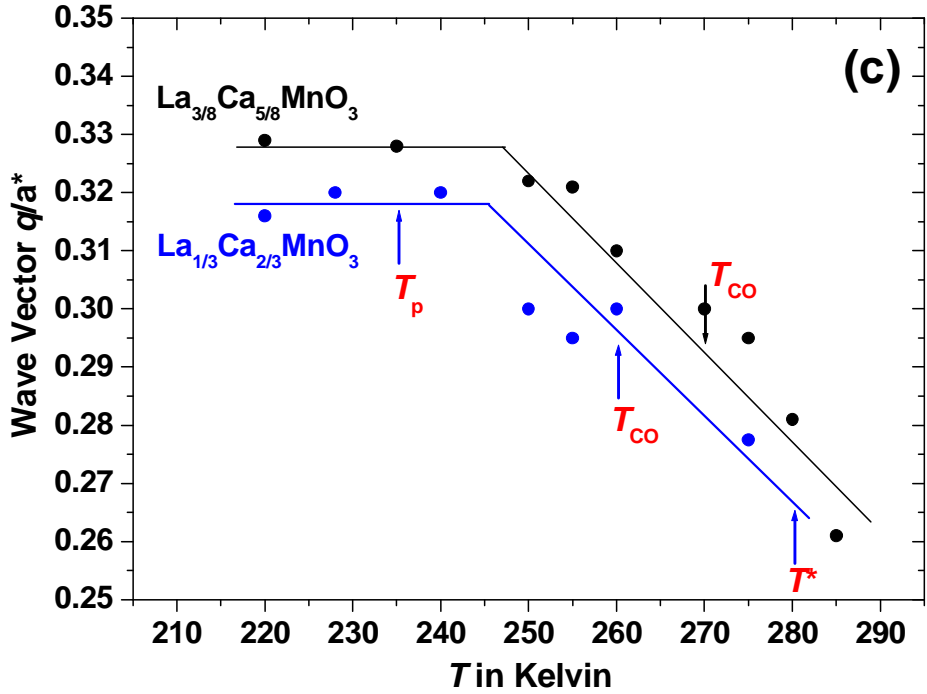

**Fig 1 (c)** (Colour Online) Charge modulation wave vector $q/a*$ in polycrystalline LCMO; for $x = 0.67$ (●) and $x = 0.625$ (○) (reproduced from [21]); $T_{CO}$ was obtained from [3]; $T*$ and $T_p$ from [14]

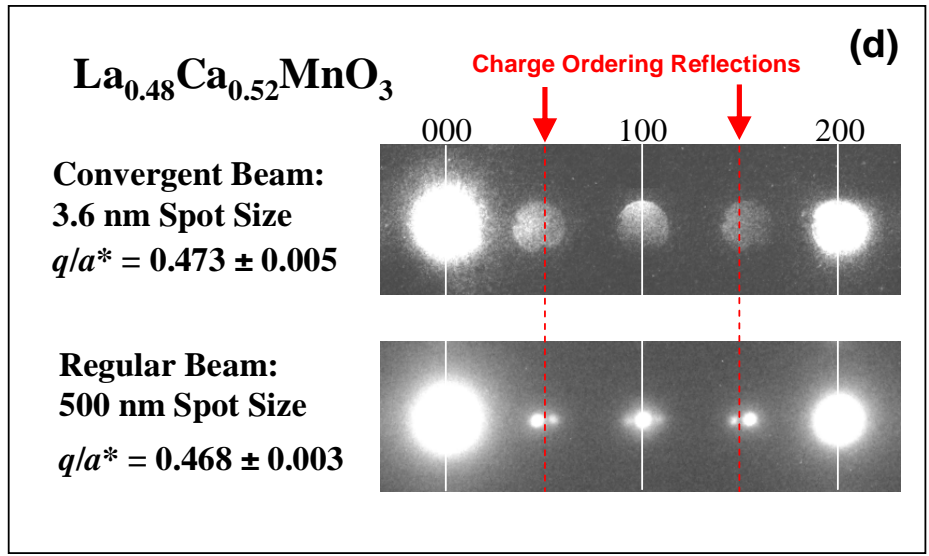

**Fig 1 (d)** (Colour Online) Electron diffraction pattern of LCMO $x = 0.52$ at 90 K showing superlattice reflections (↓) collected with a convergent electron beam of spot size 3.6 nm showing extra lattice periodicity of $q/a* = 0.473 \pm 0.005$, and with a regular spot size of 500 nm and $q/a* = 0.468 \pm 0.003$ (reproduced from [17]);

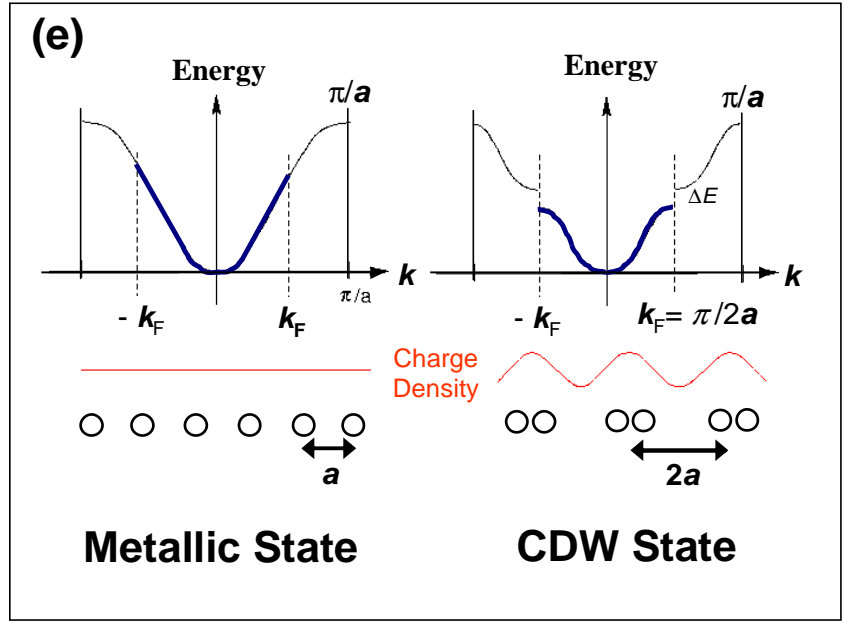

**Fig 1 (e)** (Colour Online) Metallic ($T > T_{PE}$) and CDW ($T < T_{PE}$) states in a typical CDW system: electron energy vs. electron wave vector ($k$) relationships, charge density, and lattice periodicities $a$ and $2a$, reproduced from [26]

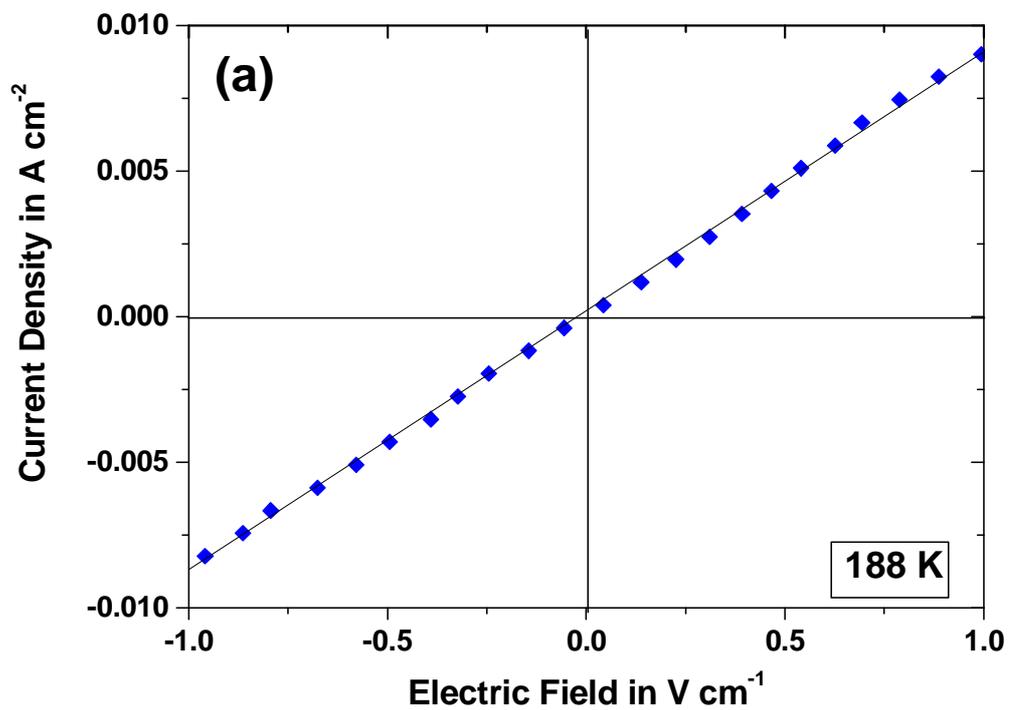

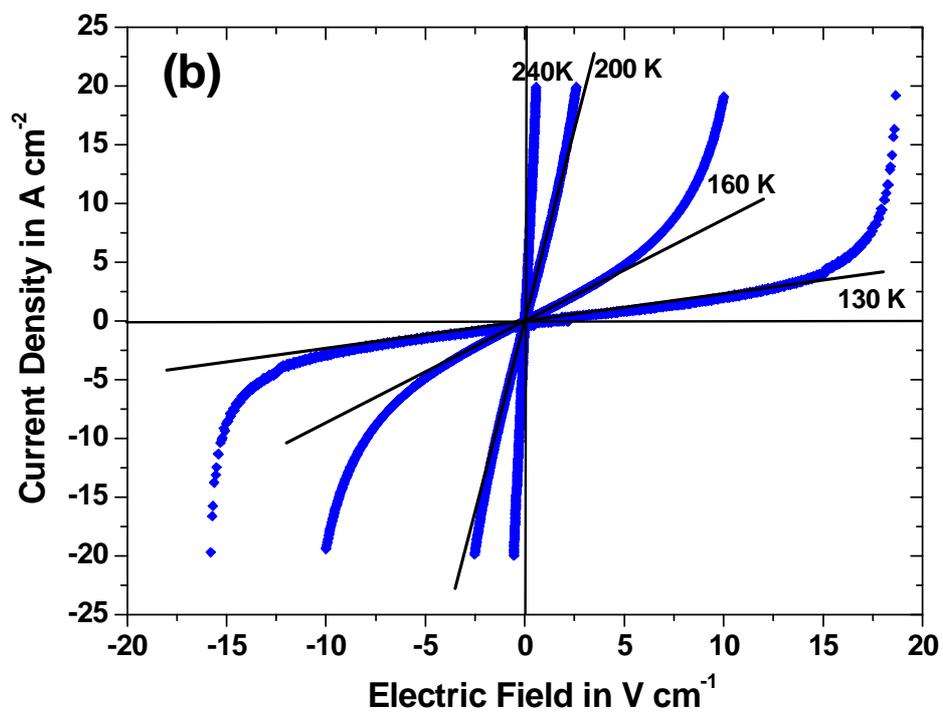

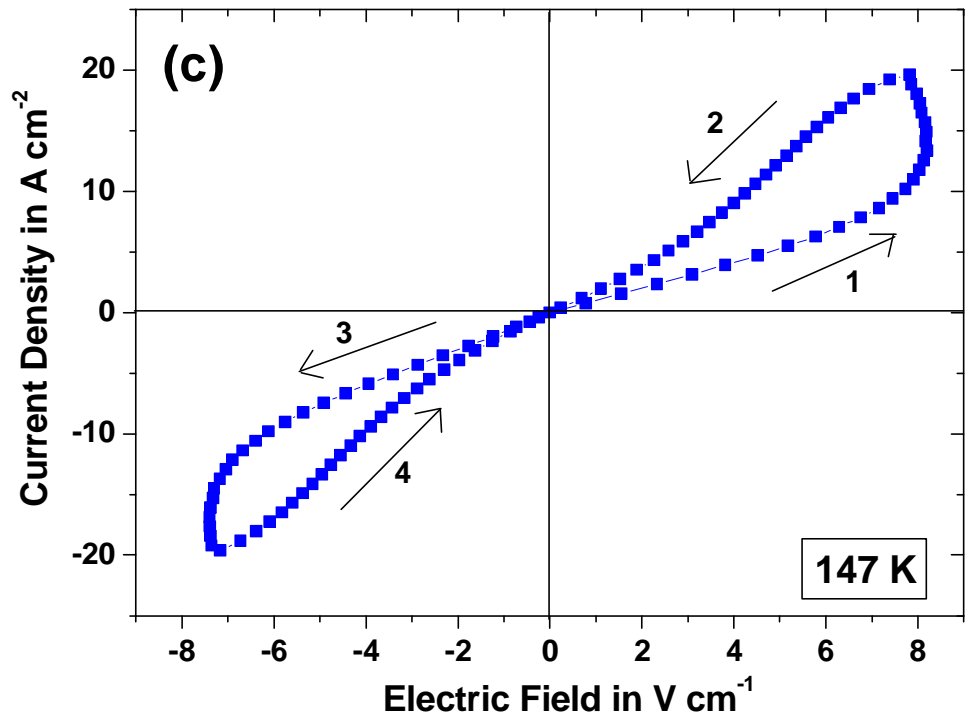

**Fig 2** (Colour Online) *I-V* characteristics of thin film and polycrystalline LCMO *x* = 0.67 **(a)** Thin film linear *I-V* along the charge modulated *a* crystal direction at the resolution limit of small applied electric fields **(b)** Polycrystals *I-V* at different temperatures; straight lines are provided as a guide to the eyes **(c)** Hysteretic *I-V* for polycrystals for an applied sine-wave amplitude voltage signal

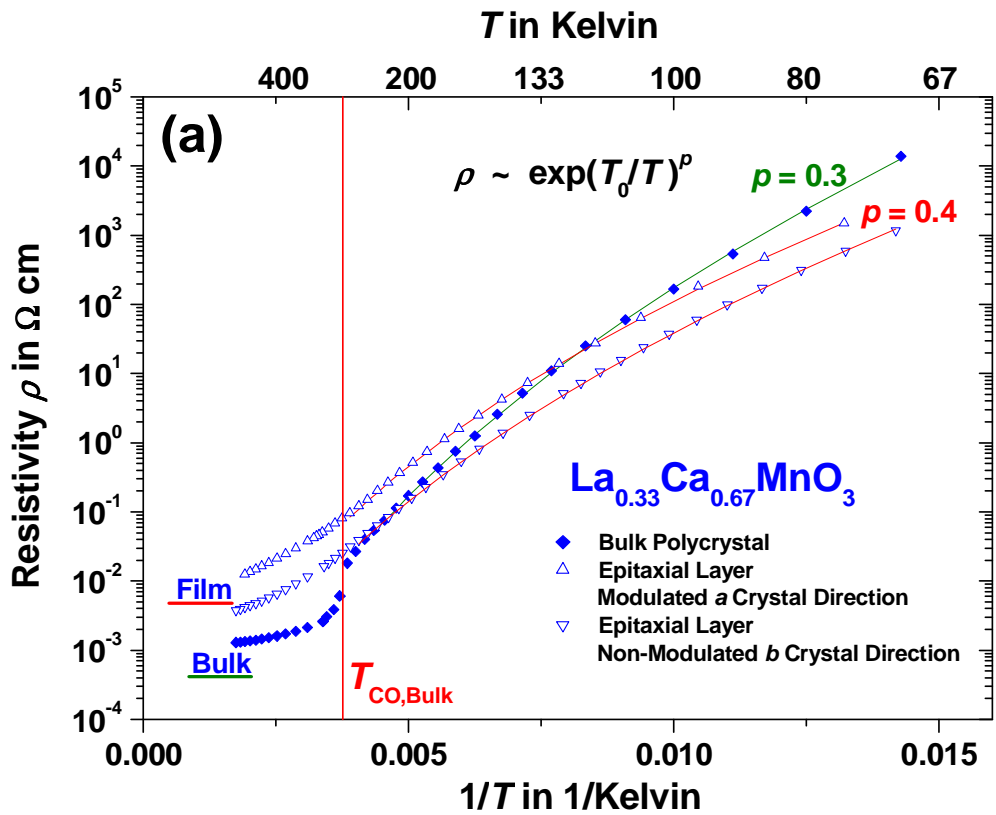

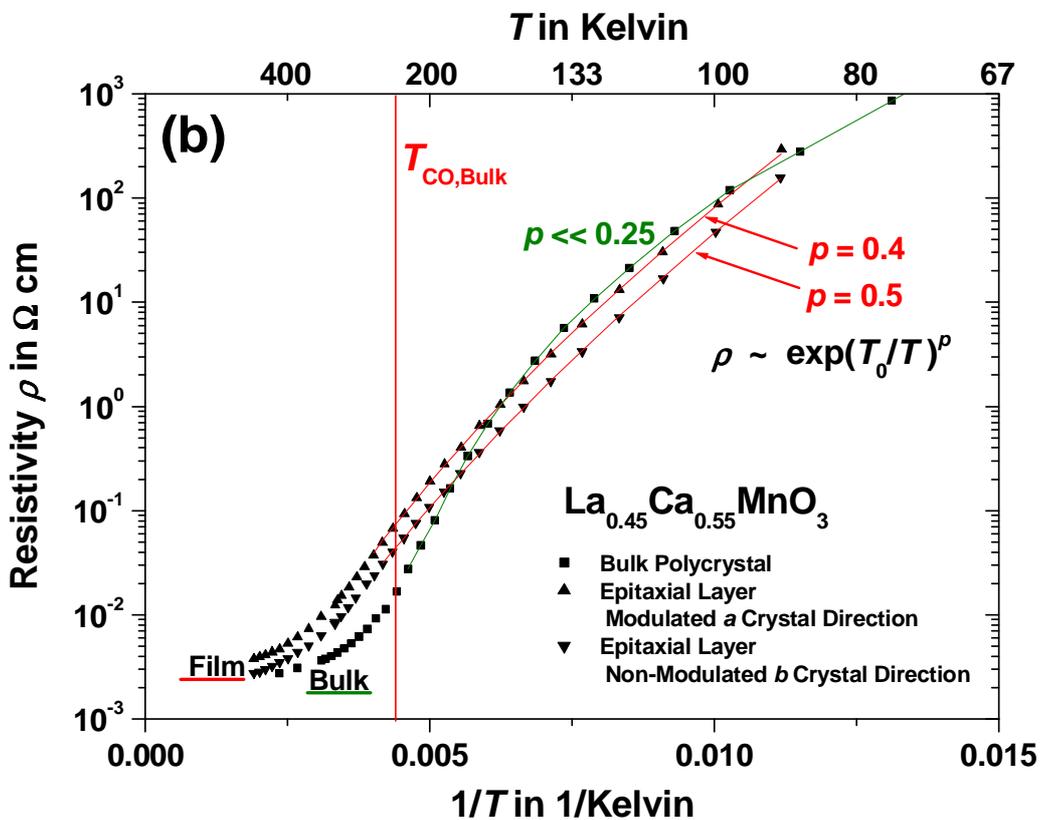

**Fig 3** (Colour Online) **(a)** ln($\rho$) vs. 1/$T$ resistivity vs. temperature characteristics for LCMO ($x = 0.67$) for polycrystals (♦), epitaxial film along the modulated ***a*** (△) and non-modulated ***b*** direction(▽); solid lines indicate temperature ranges where the data follows Variable Range Hopping behaviour (equation 1); **(b)** for LCMO ($x = 0.55$) for polycrystals (♦), epitaxial film along ***a*** (▲) and ***b*** direction (▼); solid red lines indicate temperature ranges where the data follows Variable Range Hopping behaviour (equation 1); for $x = 0.55$ polycrystals data does not follow VRH (green curve) due to $p \ll 0.25$

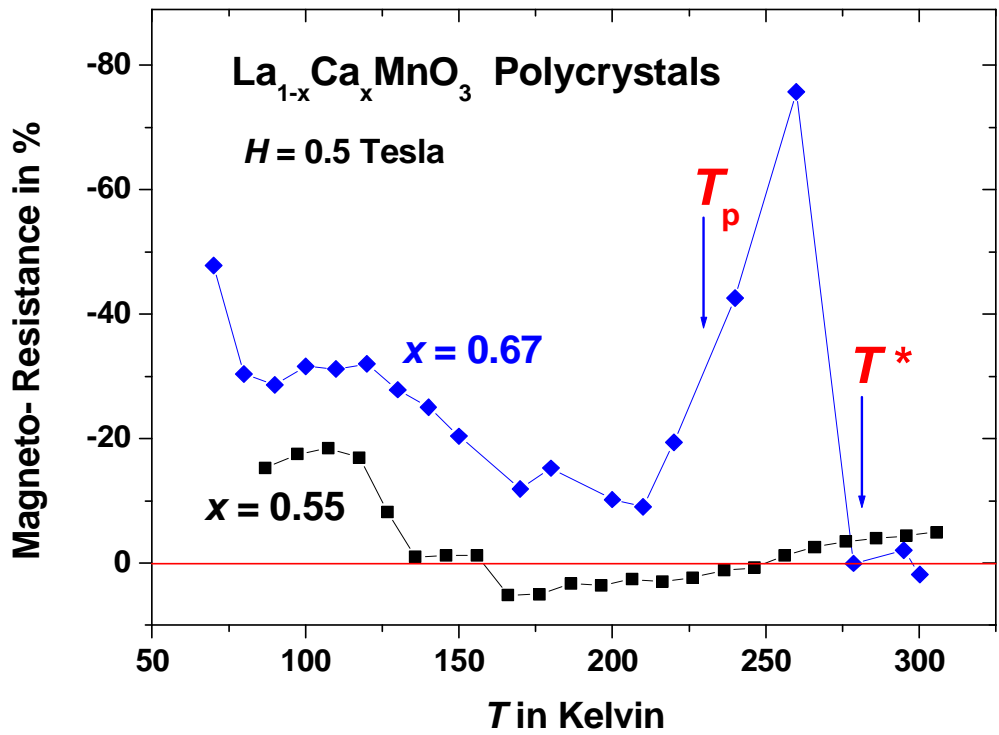

**Fig 4** Magnetoresistance (MR) at $H = 0.5$ Tesla for polycrystalline LCMO $x = 0.67$ (♦) and $x = 0.55$ (■); MR = $[\rho(0) - \rho(0.5T)]/\rho(0)$

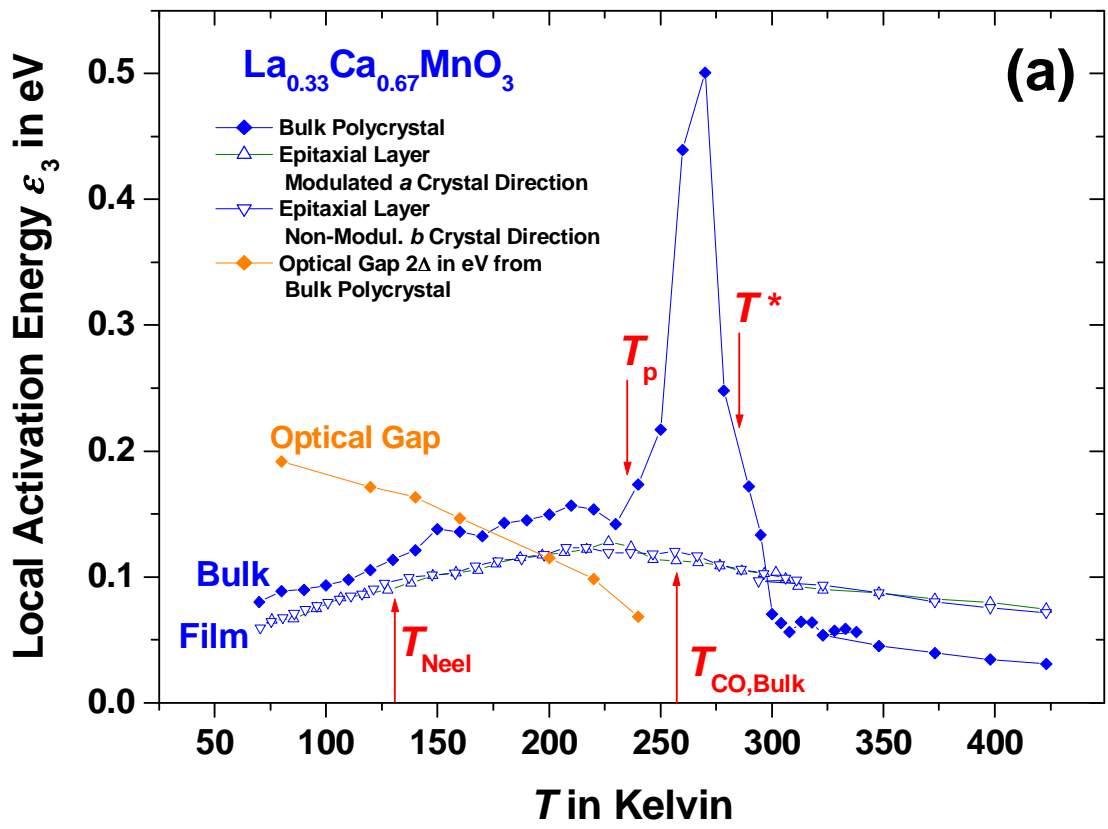
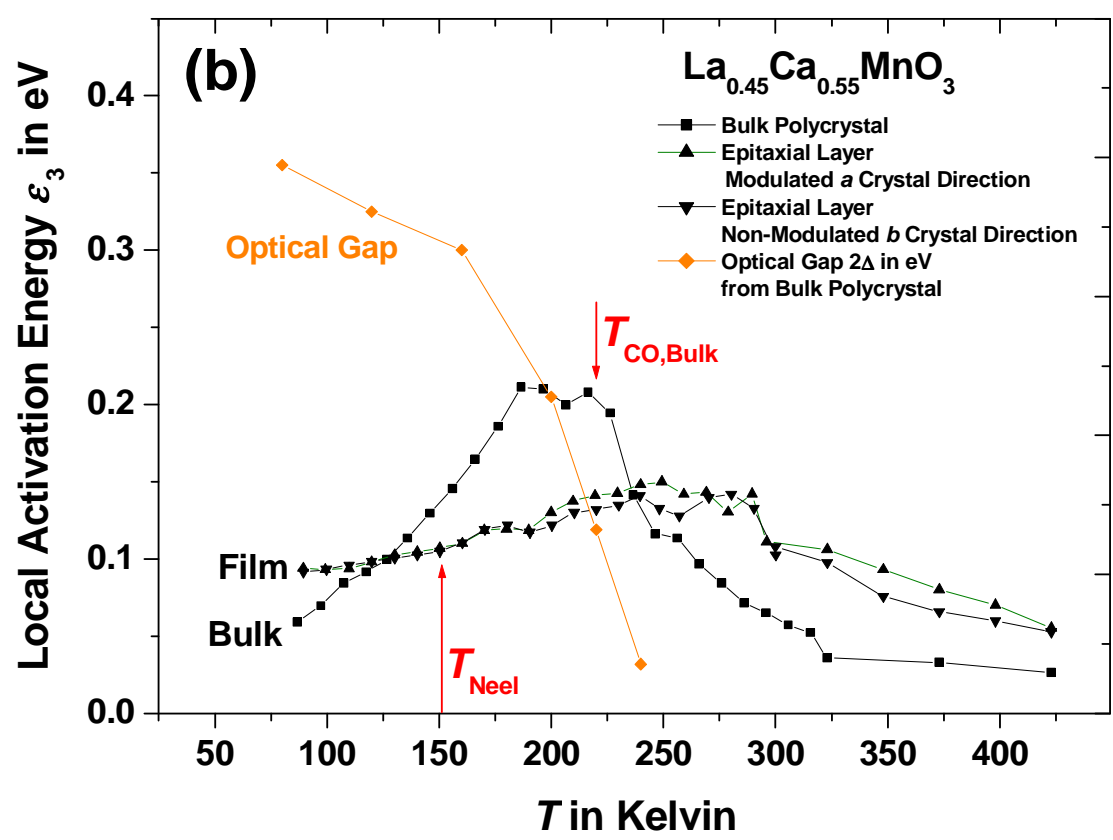

**Fig 5** (Colour Online) Activation energies $\varepsilon_3$ determined from the gradient of $\ln(\rho)$ vs. $1/T$ curves and plotted vs. $T$; **(a)** for LCMO $x = 0.67$ polycrystals (♦) and ≈ 60 nm epitaxial films for the modulated *a* (Δ) and non-modulated *b* direction (▼); optical gap $2\Delta$ (♦) (reproduced from [27]) and **(b)** for LCMO $x = 0.55$ polycrystals (■) and epitaxial layers for *a* (▲) and *b* (▼); optical gap $2\Delta$ (♦) from [27]

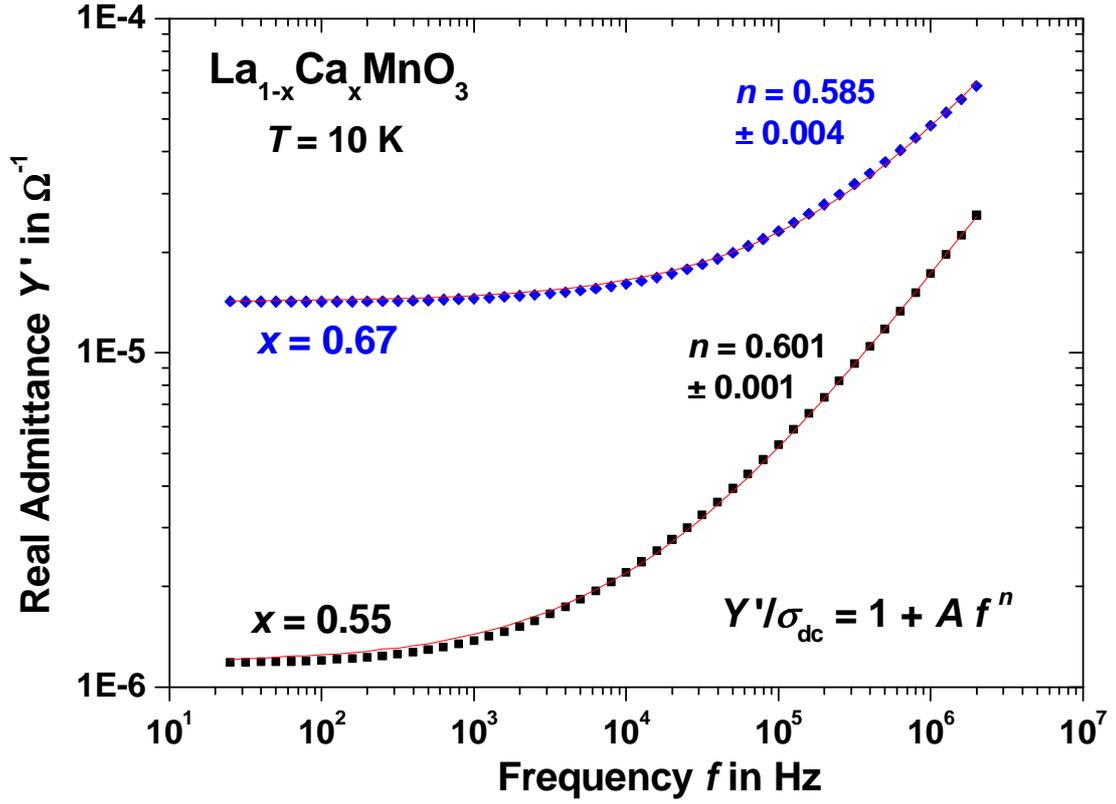

**Fig 6** Double logarithmic plots of real part of admittance vs. frequency $f$ for polycrystalline LCMO $x = 0.67$ (♦) and $x = 0.55$ (■) at 10 K; the red lines represent fits to Jonscher's law (equation 2); the critical Jonscher exponents were $n = 0.585 \pm 0.004$ and $0.601 \pm 0.001$, the purely empirical A values were $7.3 \times 10^{-4} \pm 5 \times 10^{-5}$ and $3.4 \times 10^{-3} \pm 6 \times 10^{-5}$ for $x = 0.67$ and 0.55 respectively.

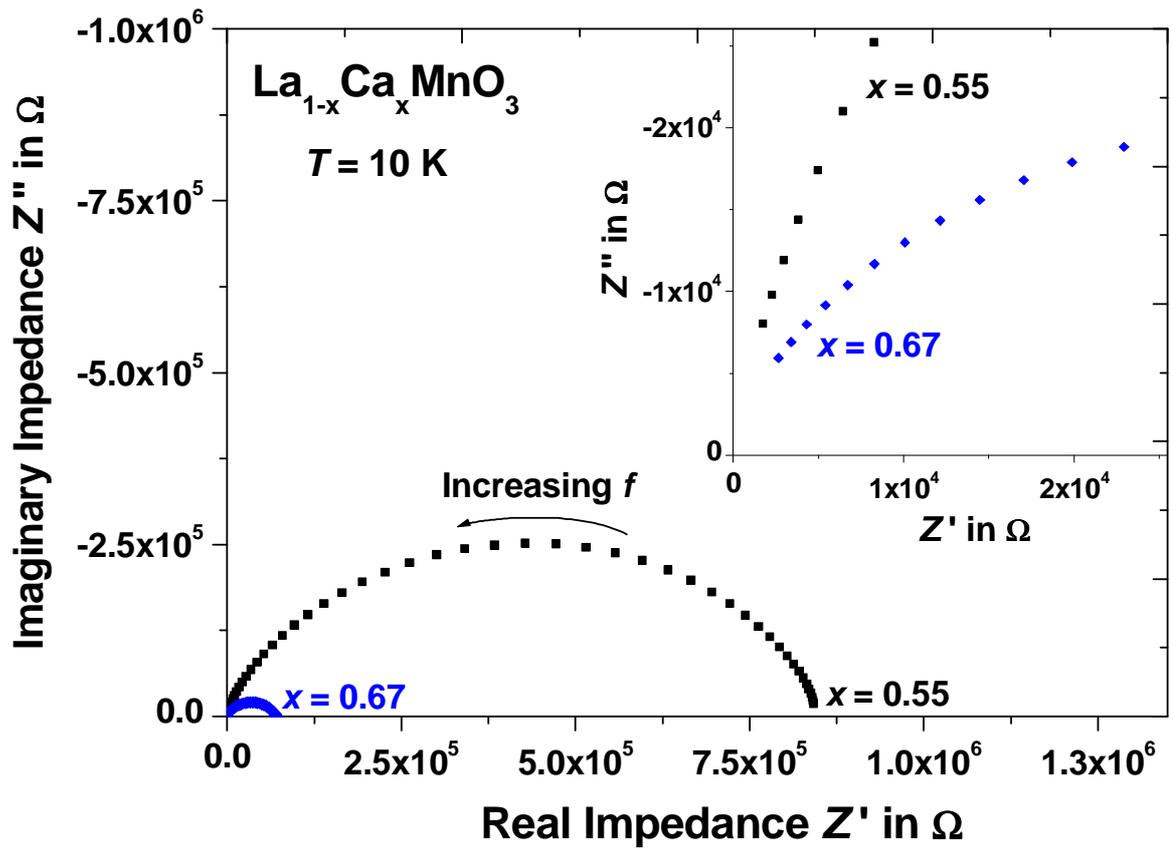

**Fig 7** Complex plane plots of real vs. imaginary parts of the impedance $Z''$ vs. $Z'$ for polycrystalline LCMO $x = 0.67$ (♦) and $x = 0.55$ (■) at 10 K; the figure inset shows a magnification of the data near the origin of the graphs

# References

*Electronic address: rainerxschmidt@googlemail.com